\documentclass[a4paper,fleqn,usenatbib]{mnras}
\usepackage{times}
\def\aj{AJ}
\def\apj{ApJ}
\def\apjl{ApJL}
\def\apjs{ApJS}
\def\ao{Appl.~Opt.}
\def\aap{A\&A}
\def\mnras{MNRAS}
\def\na{New A}
\def\prd{Phys.~Rev.~D}
\def\physrep{Phys.~Rep.}

\usepackage{newtxtext,newtxmath}

\usepackage[T1]{fontenc}
\usepackage{ae,aecompl}

\usepackage{graphicx}	
\usepackage{amsmath}	
\usepackage{amssymb}	

\title[Bubble size statistics from 21-cm tomography]{Bubble size statistics during reionization from 21-cm tomography}

\author[Giri et al.]{
Sambit K. Giri,$^{1}$\thanks{E-mail: sambit.giri@astro.su.se}
Garrelt Mellema,$^{1}$
Keri L. Dixon$^{2}$ and Ilian T. Iliev$^{2}$
\\
$^{1}$Department of Astronomy and Oskar Klein Centre, Stockholm University, AlbaNova, SE-106 91 Stockholm, Sweden\\
$^2$ Astronomy Centre, Department of Physics \& Astronomy, Pevensey III Building, University of Sussex, Falmer, Brighton, BN1 9QH, \\United Kingdom}

\date{Accepted XXX. Received YYY; in original form ZZZ}

\pubyear{2017}

\begin{document}
\label{firstpage}
\pagerange{\pageref{firstpage}--\pageref{lastpage}}
\maketitle

\begin{abstract}
The upcoming SKA1-Low radio interferometer will be sensitive enough to produce tomographic imaging data of the redshifted 21-cm signal from the Epoch of Reionization. Due to the non-Gaussian distribution of the signal, a power spectrum analysis alone will not provide a complete description of its properties. Here, we consider an additional metric which could be derived from tomographic imaging data, namely the bubble size distribution of ionized regions. We study three methods that have previously been used to characterize bubble size distributions in simulation data for the hydrogen ionization fraction - the spherical-average, mean-free-path and friends-of-friends methods - and apply them to simulated 21-cm data cubes. Our simulated data cubes have the (sensitivity-dictated) resolution expected for the SKA1-Low reionization experiment and we study the impact of both the light-cone and redshift space distortion effects. To identify ionized regions in the 21-cm data we introduce a new, self-adjusting thresholding approach based on the K-Means algorithm. We find that the fraction of ionized cells identified in this way consistently falls below the mean volume-averaged ionized fraction. From a comparison of the three bubble size methods, we conclude that all three methods are useful, but that the mean-free-path method performs best in terms of tracking the progress of reionization and separating different reionization scenarios. The light-cone effect is found to affect data spanning more than about 10~MHz in frequency ($\Delta z\sim0.5$). We find that redshift space distortions only marginally affect the bubble size distributions.

\end{abstract}




\section{Introduction}
Before the completion of hydrogen reionization, which happened around a redshift of about 6, the intergalactic medium (IGM) contained substantial amounts of neutral hydrogen. Its spin flip transition can produce an observable signal at the rest frame wavelength of 21 cm. 
This signal constitutes the most direct probe of the reionization process as it depends directly on the distribution of neutral hydrogen \citep{2006PhR...433..181F}.

Detection of this signal has therefore been one of the key science drivers for a new generation of low frequency radio interferometers, such as the Giant Metrewave Radio Telescope \citep[GMRT; e.g.][]{2011MNRAS.413.1174P}, the Low Frequency Array \citep[LOFAR; e.g.][]{2010MNRAS.405.2492H}, the Murchison Widefield Array \citep[MWA; e.g.][]{2009IEEEP..97.1497L} and the Precision Array for Probing the Epoch of Reionization \citep[PAPER; e.g.][]{2010AJ....139.1468P}. Detection of the signal is highly non-trivial due to very strong foreground signals as well as calibration challenges caused by ionospheric activity and instrumental effects. None of these arrays has yet achieved a detection, but major progress has been made and some useful upper limits have been established \citep{2015ApJ...801...51J,2017ApJ...838...65P}.

The main aim of this first generation of telescopes is to measure the (spherically averaged) power spectrum of the 21-cm signal. Extracting the power spectrum requires less signal to noise than producing images and power spectra also appear to form an excellent statistical measure of the properties of the signal as they are sensitive to both the evolution of reionization and the nature of the ionizing sources \citep[e.g.][]{2004ApJ...613....1F, 2008ApJ...680..962L, 2012MNRAS.423.2222I, 2015MNRAS.449.4246G}. Therefore, a considerable amount of effort has been invested into deriving the power spectra from models and simulations \citep[e.g.][]{2006MNRAS.372..679M, 2007MNRAS.377.1043M, 2011MNRAS.411..955M} and devising strategies to extract them reliably from the interferometer data \citep{2011PhRvD..83j3006L, 2013PhRvD..87d3005D, 2016ApJ...818..139T}.

The simulations have shown that the shape of the probability distribution function (PDF) 
of the 21-cm signal during reionization  is far from Gaussian \citep[e.g.][]{2004ApJ...613...16F, 2006MNRAS.372..679M, 2010MNRAS.406.2521I}. For this reason, the spherically averaged power spectrum does not fully describe its statistical properties. 
This non-Gaussianity can be studied using higher order statistics such as one-point skewness and kurtosis \citep[see e.g.][]{2014MNRAS.443.3090W} or the full bispectrum and trispectrum \citep[][and references therein]{2017MNRAS.472.2436W}. However, such statistics are notoriously difficult to interpret physically \citep[see e.g.][]{2016MNRAS.458.3003S}.
This forms the main motivation for the ambition to map the 21-cm signal tomographically at many different frequencies with the future Square Kilometre Array \citep[SKA;][]{2013ExA....36..235M}. Such tomographic data will show both the sizes and shapes of ionized regions as well as the density fluctuations in neutral regions and should thus give a clear view of how reionization progressed.

In order to interpret tomographic data sets in terms of, for example, the properties and distribution of the sources of ionizing photons, statistical tools for comparison to simulation results are needed. How does one characterise the tomographic data of the 21-cm signal? 
This question is not easily answered as there are no similar data sets in cosmology. The Cosmic Microwave Background is not tomographic and has a PDF which is very close to a Gaussian distribution. Galaxy redshift surveys do provide tomographic data sets, but they deal with discrete objects. However, results of those surveys can be transformed into galaxy density fields using density field estimators \citep[e.g.][]{2000A&A...363L..29S}. The algorithms developed for finding voids 
in those fields may provide some useful insights on how to deal with tomographic image data \citep[see e.g.][and references therein]{2015MNRAS.454.2228N}.

One obvious quantity in the context of reionization is the bubble size distribution (BSD), which describes how many ionized regions of a given size exist in the data. Simulations have shown that this measure describes the progression of reionization as larger and larger ionized regions appear the more reionized the Universe becomes \citep[e.g.][]{2004ApJ...613....1F, 2006MNRAS.372..679M, 2007ApJ...669..663M}. It also has appeal as a measure which appears to have a simple physical interpretation. We will therefore in this paper focus on BSDs obtained from 21-cm tomographic data sets.

BSDs have been studied before in the context of comparing simulation results. One obvious conclusion from three-dimensional simulations of reionization is that the morphology of the ionized regions is highly complex. Although cartoon versions of the process occasionally depict the ionized regions as easily identified spherical bubbles, the regions in reality have highly irregular morphologies and a complex connectivity in three dimensions. Therefore, there is no unique way to define the BSDs and different methods will give very different answers. 

For example, if one focuses entirely on connectivity, using the Friends-of-Friends (FOF) algorithm introduced in \citet{2006MNRAS.369.1625I}, one will find that well before the end of reionization most of the ionized volume becomes contained in one large connected region. The scenario is very different from the result one obtains if one focuses on the largest spherical volume which fits inside the distribution of ionized regions, a method first introduced by \citet{2007ApJ...654...12Z} and known as the Spherical Average method (SPA). Yet another result is obtained if one finds the distribution of the distances to the edge of an ionized region from a large collection of random points and random directions, a method developed by \citet{2007ApJ...669..663M}, also known as the mean free path (MFP) method.

Each of these methods have their own definition of bubble size and measure essentially different things. In order to compare inferences from different methods, one should be aware of this difference. \citet{2011MNRAS.413.1353F} and recently \citet{2016MNRAS.461.3361L} have analysed these methods in the context of characterising the differences between the ionization fraction results of different simulations. Those authors have pointed out the various characteristics, advantages and disadvantages of the different methods. 

Here we will take these methods and instead apply them to (simulated) 21-cm data. We will address the impact of resolution and the fact that 3D tomographic data will be in the form of light-cones where the signal evolves along the frequency axis. In addition, we will consider the effect of redshift space distortions \citep[RSDs;][]{2013MNRAS.435..460J,2016MNRAS.456...66J} on the BSDs. For now, we will not take into account other observational effects such as noise and calibration errors. We will restrict ourselves to the three well established methods mentioned above, FOF, SPA and MFP, and not consider methods which have recently been proposed such as the watershed algorithm \citep{2016MNRAS.461.3361L} or granulometry \citep{2017MNRAS.471.1936K}. The basic questions we are trying to answer is how these methods can be applied to observed 21-cm tomographic data and how well they perform in this context.

The paper is structured as follows. In the next section, we describe the quantity from which we will derive the BSDs, namely the redshifted 21-cm signal. Section 3 provides an overview of the size determination methods and the procedures to use them on 21-cm measurements. Section 4 gives a description of the simulations used in this study as well as how we generate the mock observations. Section 5 presents the results of our study. The sixth section contains the discussion along with the summary.

\section{Redshifted 21-cm signal}\label{sec:21cm_signal}
The neutral hydrogen 21-cm line will be a very powerful tool to study the EoR \citep{1997ApJ...475..429M}. It is a hyper-fine line of wavelength 21 cm, caused by the ground state spin-flip transition in the atom's electron-proton configuration. The neutral hydrogen atoms in the Intergalactic Medium (IGM) during reionization could be observed through the redshifted 21-cm signal using low frequency radio telescopes. The signal is seen against the Cosmic Microwave Background (CMB) and the measured differential brightness temperature can be written as \citep[e.g.,][]{2013ExA....36..235M}:
\begin{eqnarray}
\delta T_\mathrm{b} \approx 27 x_\mathrm{HI} (1 + \delta)\left( \frac{1+z}{10} \right)^\frac{1}{2}
\left( 1 -\frac{T_\mathrm{CMB}(z)}{T_\mathrm{s}} \right)\nonumber\\
\left(\frac{\Omega_\mathrm{b}}{0.044}\frac{h}{0.7}\right)
\left(\frac{\Omega_\mathrm{m}}{0.27} \right)^{-\frac{1}{2}} 
\left(\frac{1-Y_\mathrm{p}}{1-0.248}\right)
\mathrm{mK}\,,
\label{eq:dTb}
\end{eqnarray}
where $x_\mathrm{HI}$ and $\delta$ are the neutral hydrogen fraction and the density fluctuation respectively, $T_\mathrm{s}$ is the spin temperature or the excitation temperature of the distribution of the two states of hydrogen. $T_\mathrm{CMB}(z)$ is CMB temperature at redshift $z$ and $Y_\mathrm{p}$ is the primordial helium abundance.

The above equation shows that the 21-cm signal would not be observed when $T_\mathrm{s}$ is fully coupled to the CMB temperature $T_\mathrm{CMB}$. During EoR, the spin temperature is expected to approach the gas temperature due to the Wouthuysen-Field effect \citep{1997ApJ...475..429M} and the 21-cm signal would be visible with CMB as the background. When the gas temperature is below $T_\mathrm{CMB}$, the signal is seen in absorption and when it is above, it is seen in emission. For the case $T_\mathrm{s} \gg T_\mathrm{CMB}$, typical for the later stages of reionization, $1 -\frac{T_\mathrm{CMB}(z)}{T_\mathrm{s}} \rightarrow 1$ and the signal becomes independent of the value of $T_\mathrm{s}$. Throughout this paper we will use this high spin temperature limit.

The signal is observed at a frequency given by 
\begin{equation}
\nu_\mathrm{obs} = \frac{\nu_0}{1 + z_\mathrm{obs}}=
\frac{\nu_0(1-v_\|/c)}{1 + z}\,,
\label{eq:rsd_frequency}
\end{equation}
where $\nu_0=1.42$~GHz is the frequency of the transition and $z_\mathrm{obs}$ is the observed redshift which is different from the cosmological redshift $z$ due to line of sight component of the peculiar motions in the intergalactic gas, $v_\|$. This causes a distortion of the signal when observed in redshift space. See \citet{2012MNRAS.422..926M} for a comprehensive description of this RSD effect.

Observations will produce a three-dimensional data set $\delta T_\mathrm{b}(\boldsymbol{\theta},\nu_\mathrm{obs})$ where $\boldsymbol{\theta}$ indicates a position in the sky. Since $\nu_\mathrm{obs}$ depends on the cosmological redshift $z$, this data set will cover a range of look back times. The fact that the signal at different frequencies originates from different look back times is known as the light-cone effect and the data set $\delta T_\mathrm{b}(\boldsymbol{\theta},\nu_\mathrm{obs})$ is referred to as a light-cone.

\section{Bubble size statistics methods}
\label{sec:BS}
 
In this section, we first provide a brief description of all the size determination methods which we explore in this paper. For a more detailed description of the methods, we encourage the readers to consult the original papers as well as \citet{2011MNRAS.413.1353F} and \citet{2016MNRAS.461.3361L}. After this we describe the method we use to identify ionized regions in 21-cm observations.

\subsection{Methods}
\label{sec:BS_methods}
\subsubsection{Mean-free-path (MFP)}
This method was introduced in \citet{2007ApJ...669..663M} and is based on Monte-Carlo inference. It selects a random ionized location and casts a ray in a random direction. The ray is followed until a stopping criteria is met at which point the length of the ray is recorded. This process is repeated numerous times, in our case $10^7$ times. The final result is a histogram of ray length values. Different stopping criteria can be defined. When applied to a binary ionization fraction field in which cells are labelled as either ionized or neutral, the ray is stopped when it reaches the first neutral cell. This is how we will use this method.\footnote{If the MFP method is applied to an {\sc HI} density field, one can also calculate the optical depth for hydrogen ionizing photons $\tau$ along the ray and use a limit on $\tau$ as a stopping criterion, see \citet{2008MNRAS.391...63I} for an example of this approach.} The MFP method derives its name from the fact that the ray traced corresponds to the {\lq mean free path\rq} of an ionizing photon, given that the ionized region is typically highly ionized and neutral region nearly completely neutral. 

In the simplest implementation, semi-numerical methods, such as 21cmFAST \citep{2011MNRAS.411..955M}, directly produce binary fields. Fully numerical methods such as C$^2$-Ray produce continuous values between 0 and 1 for the ionization fraction $x_\mathrm{HII}$. This means that a certain threshold value has to be chosen to convert this continuous field to a binary field. Often $x_\mathrm{HII}=0.5$ is chosen. As shown in \citet{2011MNRAS.413.1353F} the MFP-BSD is sensitive to the precise choice of this threshold value. 

A BSD method is diffusive if applying it to a collection of non-overlapping bubbles of radius $R_0$ will not yield the correct BSD which is $\delta(R-R_0)$, but rather a distribution stretching over a range of bubble sizes. For the MFP method ray lengths can vary from 0 to $2R_0$ and therefore it is diffusive. A BSD method is biased if the peak of the distribution is not at $R_0$ but at a different size. \citet{2016MNRAS.461.3361L} showed that the MFP-BSD peaks very close to $R_0$ and classified the method as unbiased.

\subsubsection{Spherical-average (SPA)}
The spherical-average method was proposed in \citet{2007ApJ...654...12Z} and used to compare the analytic calculation of the BSD from excursion set theory with results from radiative transfer simulations. For each ionized location, this method finds the largest sphere around it for which the average ionization fraction is above some chosen threshold value. The final result point is a histogram of radius values which is meant to represent the BSD. Since this method evaluates the average ionization fraction in a certain region it requires a threshold value even in the case of a binary ionization fraction field. Since we want our bubbles to be mostly ionized, threshold values close to 1 are chosen. We will use 0.9.

As pointed out by \citet{2011MNRAS.413.1353F} and \citet{2016MNRAS.461.3361L}, the SPA method is both diffusive and strongly biased. For a collection of non-overlapping bubbles of radius $R_0$, the bubble sizes range from 0 to $R_0$ and the peak of the distribution is found close to $R_0/3$. Note that \citet{2016MNRAS.461.3361L} refer to this method as the Distance Transform (DT).

\subsubsection{Friends-of-friends (FOF)}
The friends-of-friends method 
is based on the idea of hierarchical clustering used in the field of data mining and statistics \citep{ivezic2014statistics}. The linkages between data points are used to find clusters. If a data point is within a (chosen) linking length of any of the points in a cluster, it becomes part of that cluster. This method is extensively used for cluster analysis in $N$-body simulations \citep[e.g.,][]{1982ApJ...259..449P,1985ApJ...292..371D}. 
\citet{2006MNRAS.369.1625I} first used FOF to analyse gridded $x_\mathrm{HII}$ data from an EoR simulation by linking any neighbouring cells which have been labelled as ionized. For each cluster, the volume is calculated and the final result is a histogram of cluster volumes. \citet{2016MNRAS.457.1813F} pointed out that this approach is the same as the Hoshen-Kopelman algorithm \citep{hoshen1976percolation}. The typical histogram, particularly in the middle and late stages of reionization shows a bimodal distribution in which one large, percolated cluster contains most of the ionized points and a collection of much smaller isolated clusters contain the remainder. \citet{2011MNRAS.413.1353F} and \citet{2016MNRAS.457.1813F} analysed this property further. The latter authors showed how this dominant cluster grows as reionization progresses. They also found this feature to be universal and due to the fact that reionization is percolation process. 

Just as the MFP method, FOF does not require a threshold when applied to a binary field. However, as explained above the generation of a binary field from a continuous field does require the choice of a threshold. The FOF method is neither diffusive nor biased, but it measures volumes of topologically-connected ionized regions and not radii.

\subsection{Binary field creation}
\label{sec:ThS}

\begin{figure*}
  \includegraphics[width=0.9\textwidth]{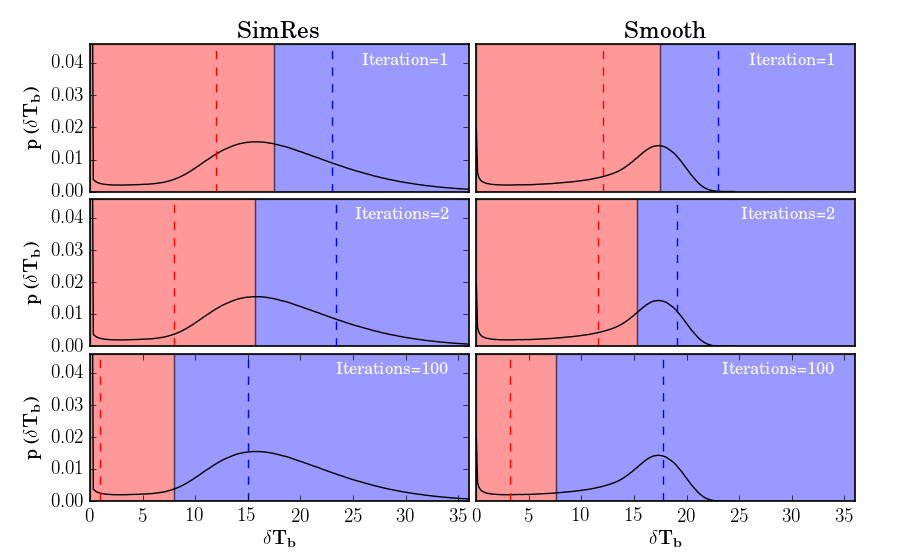}
  \caption{Illustration of the steps followed by K-Means to determine the threshold from the PDF has been shown. \textit{Row 1:} Two random points (cluster centres) are chosen and all the other points are linked to these values based on their distance. The dashed lines show the cluster centres and the colour of the shaded region shows linkage. \textit{Row 2:} The mean of all the points present in each shaded region of {\lq row 1\rq} gives the new cluster centres and the same linking process is repeated. \textit{Row 3:} After many iterations of the same process, the cluster centres cease to shift, and the algorithm has converged.}
  \label{fig:kmeans_cart}
\end{figure*}
%
%
All the methods discussed above have been previously applied to ionization fraction fields. Depending on the origin of the field, these were either binary fields or continuous fields transformed into binary ones through the choice of an appropriate threshold value, for example $x_\mathrm{HII}=0.5$. However, we now want to apply these methods to data sets containing the observed value of $\delta T_\mathrm{b}$. We therefore require a method to transform $\delta T_\mathrm{b}$ into a binary field of neutral and ionized regions. It is difficult to define a fixed threshold value to achieve this as $\delta T_\mathrm{b}$ depends on both density and ionization fraction variations, as well as on redshift. Fully ionized regions will of course have $\delta T_\mathrm{b}=0$. However, {\it interferometric} tomographic data, due to the lack of baselines of length zero, do not allow the measurement of the absolute value of $\delta T_\mathrm{b}$. Furthermore, the observations will not resolve scales below several comoving Mpc and will therefore most likely not resolve the ionization fronts, thus further complicating the choice of a threshold.

The problem of dividing an image or three-dimensional data set into different regions is called segmentation and in the field of image processing, many methods exist which use the data itself to achieve this. One obvious way for the case of 
the 21-cm signal is to consider its PDF. Previous works \citep[see e.g.][]{2010MNRAS.406.2521I} have shown this PDF to be bimodal. 
This property allows automatic selection of a threshold value in the $\delta T_\mathrm{b}$ data to label regions as either ionized or neutral and hence create the binary field.

To automatically select an appropriate threshold value, we use the K-Means clustering algorithm \citep[e.g.,][]{1979HartiganWong, kanungo2002efficient}. 
K-Means is an unsupervised clustering method that finds clusters in large datasets \citep[e.g.][]{2013ApJ...763...50S}. A bimodal PDF has the data points clustered at the two peak values of the modes. K-Means then finds these two clusters and puts a threshold in between them.
The method starts with choosing two random values in the range of values present in the data cube. These are the initial guess for the cluster centres. All other values are connected to one of these points to form two clusters. Next the centroids of these two clusters are found. Using these calculated centroids as the new centre of the clusters, the clustering of the points is recalculated. This process is repeated until the calculated centroids overlap with the cluster centres.

Fig.~\ref{fig:kmeans_cart} illustrates our threshold selection process pictorially. Here, we start with guesses for the initial cluster centres that are far away from their final position. The cluster centres converge to the required positions when the K-Means algorithm is allowed to run a sufficient number of iterations. It has been shown that the algorithm always converges to the solution in a finite time \citep[e.g.][]{Bottou95convergenceproperties,kanungo2002efficient}, which makes it an apt choice for our case. We note that unlike some global thresholding methods, like Otsu's method \citep{otsu1979threshold}, K-Means does not actually construct a PDF but works directly on the data points. However, for the one-dimensional case considered here (values of the 21-cm signal) both methods produce similar results \citep{liu2009otsu}. Both the one-dimensional version of K-Means and Otsu's method become unreliable when the PDF does not possess a clear bimodality. 

We will evaluate the performance of K-Means below in Section~\ref{sec:global_ionfrac}. As the continuous ionization fraction field also displays a bimodal PDF \citep[see e.g. Fig~36 in][]{2006MNRAS.371.1057I}, K-Means can also be used when analysing such simulation results, as we will show below. We like to point out that there exist other algorithms to produce binary ionization fields from 21-cm data, which do not necessarily rely on the PDF. We will explore such other segmentation approaches in a future paper (Giri et al., in preparation).

\begin{figure*}
  \includegraphics[width=0.9\textwidth]{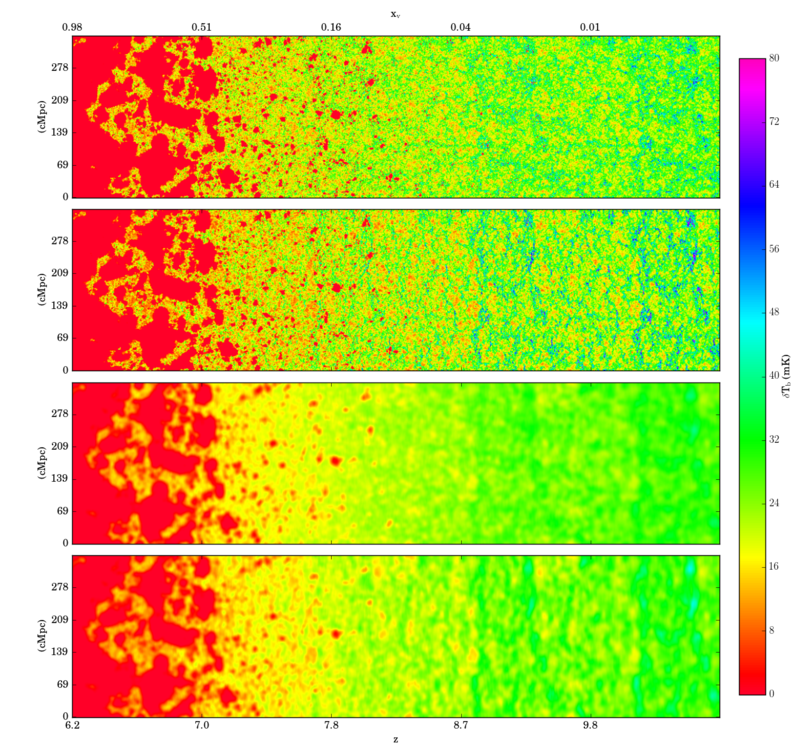}
  \caption{A cut along the light-cone from our fiducial simulation. The top two panels show the results at the resolution of the simulation without and with RSD effects respectively. Similarly, the bottom two panels show those light-cones after smoothing with a Gaussian beam corresponding to the resolution of a radio interferometer with a maximum baseline of 2 km. In the frequency direction the light-cone the data has been smoothed with a top-hat kernel of the same width as the FWHM of the corresponding angular smoothing kernel.
  The colour-bar shows the absolute value of the differential brightness temperature $\delta \mathrm{T_b}$. The vertical axes of the light-cone slice gives the length in comoving units. The horizontal axis in the bottom panel shows the redshift values whereas the top panel indicates the corresponding mean ionization fraction.}
  \label{fig:LC_SmoothLC}
\end{figure*}

\subsection{BSD curves}\label{sec:BS_curve}
After running the various BSD algorithms, we obtain the number of bubbles within a given size range. 
For both MFP and SPA, we will plot the following quantity against size $R$,
\begin{equation}
P(R) = R\frac{\mathrm{d}n}{\mathrm{d}R} = \frac{\mathrm{d}n}{\mathrm{d}\log(R)}\,,
\end{equation}
where the latter expression shows that $P(R)$ is equivalent to using logarithmic binning of $R$. This quantity gives the fraction of bubbles that fall in a given size range, which means that it does not provide information on {\it how many} bubbles have been identified.

Plotting the equivalent $P(V)$ curves for the volumes determined from the FOF method has the undesirable effect that the single largest cluster, which contains most of the ionized volume, becomes unnoticeable compared to the many small regions. To make the contribution of this largest cluster more prominent, 
we instead plot $V/V_\mathrm{ionized}P(V)$ for the sizes determined from FOF, where $V_\mathrm{ionized}$ is the total volume of ionized regions. This quantity therefore shows what fraction of the total ionized volume is contained in regions of a given volume.

\section{Simulated 21-cm signal}
\subsection{Numerical simulation}
\label{num_sim}

\begin{table*}
	\centering
	\caption{Parameters of the reionization simulations that is used in this study. The labels used are the same as the ones used in \citet{2016MNRAS.456.3011D}, where these simulations were first introduced.}
	\label{tab:sim_model}
	\begin{tabular}{lcccccc} 
		\hline
		Label & Box size (Mpc) & $g_{\gamma}$ (HMACH) & $g_{\gamma}$ (LMACH) & $g_{\gamma}$ (LMACH)$_\mathrm{supp}$ & Mesh & $\tau$\\
		\hline
		LB1 (fiducial) & 349 & 1.7 & 0 & 0 & $250^3$ & 0.049\\
		LB3 & 349 & 1.7 & 7.1 & 1.7 & $250^3$ & 0.068\\
		LB4 & 349 & 1.7 & 1.7 & eq 4 of \citet{2016MNRAS.456.3011D}  & $250^3$ & 0.057\\
		\hline
	\end{tabular}
\end{table*}

To investigate the use of the different BSD algorithms on 21-cm signal data cubes, we use the results from large-scale fully numerical reionization simulations
. The details of our simulation methodology have been discussed in previous papers \citep{2006MNRAS.372..679M,2006MNRAS.369.1625I,2012MNRAS.424.1877D}. In short, we follow the
evolution of matter with an $N$-body simulation using the \textsc{CUBEP$^3$M} code \citep{2013MNRAS.436..540H}. We then postprocess the results with a radiative transfer simulation using the \textsc{C$^2$-RAY} code \citep{2006NewA...11..374M} where we assign an ionizing luminosity based on physically-motivated models to the haloes found in the $N$-body simulation.

The specific simulations we have used follow reionization in a comoving volume of $(349\mathrm{Mpc})^3$. We assume a $\Lambda$CDM universe with cosmological parameters  $\Omega_\mathrm{m}$=0.27, $\Omega_\mathrm{k}$=0, $\Omega_\mathrm{b}$=0.044, $h=0.7$, $n=0.96$, $\sigma_8$=0.8 and $Y_\mathrm{p}=0.248$, consistent with the \textit{Wilkinson Microwave Anisotropy Probe} (WMAP) \citep{2011ApJS..192...18K} and Planck \citep{2016A&A...594A..13P} results. We will use three different assumptions for the source properties, labelled LB1, LB3 and LB4. These simulations were described and studied in detail in \citet{2016MNRAS.456.3011D}. A summary of the source parameters used for those simulations is given in Table~\ref{tab:sim_model}. LB1 is our fiducial case. In this case the only active sources are located in dark matter haloes of masses larger than $10^9$~M$_\odot$ (high mass atomically cooling haloes or HMACHs). These haloes release ionizing photons at a rate of 1.7 photons per baryon every $10^7$~years. Simulation LB3 uses additional low-mass sources with halo masses between $10^8$ and $10^9$~M$_\odot$ (low mass atomically cooling haloes or LMACHs) with an ionizing photon rate of 7.1 per baryon every $10^7$~years. These haloes are assumed to be subject to radiative feedback and their ionizing photon rates drops to 1.7 photons per baryon every $10^7$~years once they are located inside an ionized region. In the LB4 case, the same low-mass sources are used, but the radiative feedback is implemented by a mass-dependent suppression factor in ionized regions, as described in \citet{2016MNRAS.456.3011D}. Apart from the simulation parameters Table~\ref{tab:sim_model} also lists the value for the electron scattering optical depth derived from these simulations. The values are all consistent within 1-$\sigma$ with the measurements by the Planck satellite \citep{2016A&A...596A.107P,2016A&A...596A.108P}.

We construct $\delta T_\mathrm{b}(\mathbf{x},z)$, the differential brightness temperature, originating from a given location at a given time corresponding to cosmological redshift $z$ using equation~(\ref{eq:dTb}). The ionization fractions $x_\mathrm{HII}$ are produced by the radiative transfer simulation, while the density fluctuations $\delta$ are taken from the results of the $N$-body simulation. Those density fluctuations have been smoothed and gridded into the radiative transfer mesh. Since all values in this three-dimensional data set correspond to the same cosmological redshift $z$, we refer to it as coeval cubes (CC) to distinguish it from the light-cone (LC) data set discussed next.

\subsection{Light-cone construction}

As explained in Section~\ref{sec:21cm_signal}, the data set observed by a radio interferometer is a light-cone in which the images at different frequencies correspond to different signals originating from different redshifts and which are in addition distorted due to peculiar motions in the intergalactic gas (i.e.\ the RSDs). We construct light-cone data from the coeval simulation data using the procedure described in \citet{2006MNRAS.372..679M} and \citet{2012MNRAS.424.1877D}. The neutral fraction light-cone and the density light-cone are constructed separately and then are used to construct a 21-cm signal light-cone using equation~(\ref{eq:dTb}). To be able to study the impact of the RSD we produce two types of light-cone data, one without RSD for which $z_\mathrm{obs}=z$ and one with RSD, using equation~(\ref{eq:rsd_frequency}). We account for the RSD in the light-cone using the MM-RRM scheme explained in \citet{2012MNRAS.422..926M}. {The top two panels of Fig.~\ref{fig:LC_SmoothLC} show cuts along a non-distorted and a redshift space distorted light-cone from our fiducial simulation.}

\subsection{Telescope resolution}
The typical Full Width Half Maximum (FWHM) of the point spread function of an interferometer is given by \citep[e.g.][]{rohlfs2013tools},
\begin{equation}
\label{eq:res}
\Delta \theta = \frac{\lambda}{B} \,.
\end{equation}
In the above equation, $\lambda$ is redshifted value of $\lambda_{21}$ (i.e.\ $21.1(1+z)$ cm) and $B$ represents the maximum baseline length used for producing the images. Unless otherwise specified, we use the planned maximum baseline of the core of SKA1-Low, which is 2 km. We will refer to this as SKA1-Low resolution although the actual resolution may be slightly different.

To mimic the response of SKA1-Low, we smooth the light-cone with a Gaussian kernel of FWHM $\Delta\theta$ in the angular direction. This FWHM is frequency dependent as the resolution of the radio telescope decreases as we go to higher redshift. 
In the frequency direction of the light-cone, we smooth the data with a top-hat kernel of the same width as the FWHM of the corresponding angular smoothing kernel. {The two lower panels of Fig.~\ref{fig:LC_SmoothLC} illustrates how this smoothing affects the simulated signals for both the non-distortied and the redshift space distorted case.} 

\section{Results}
\subsection{Global ionization fraction}
\label{sec:global_ionfrac}
\begin{figure}
  \centering
  \includegraphics[width=0.5\textwidth]{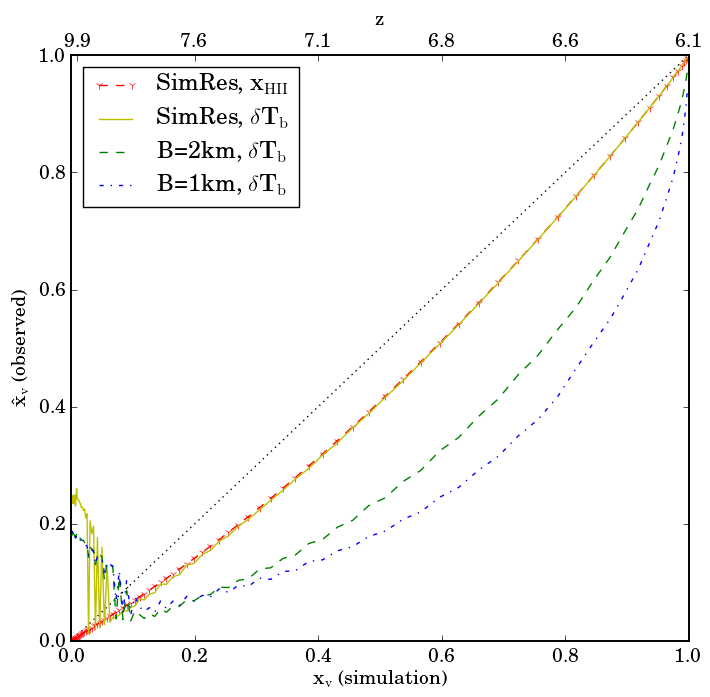}
  \caption{The fraction of ionized cells $\hat{x}_\mathrm{v}$ identified by the K-Means algorithm against the mean volume-weighted ionization fraction $x_\mathrm{v}$ from the simulation. K-Means was used to produce binary fields from following data sets: the ionized fraction $x_\mathrm{HII}$ at the resolution of the simulation and the differential brightness temperature at three different resolutions: that of the simulation, and those corresponding to maximum baselines of 2 km and 1 km.}
  \label{fig:x_obs_kmeans}
\end{figure}

After segmenting a tomographic 21-cm data set into a binary ionization fraction 
field, the first quantity to consider is the global ionized fraction by volume, $x_\mathrm{v}$, or in other words what fraction of space is contained in ionized regions. This quantity is easily calculated from simulation results but for the observations will depend on the chosen segmentation as well as the resolution. In Fig.~\ref{fig:x_obs_kmeans}, we show the measured global ionization fraction $\hat{x}_\mathrm{v}$ against the actual one  $x_\mathrm{v}$ for our fiducial simulation for the entire reionization history. We consider four different binary fields. The first two were generated from the ionized fraction and $\delta T_\mathrm{b}$ fields at the resolution of the simulation. The other two were obtained from $\delta T_\mathrm{b}$ fields where we reduced the resolution to the SKA1-Low case and twice worse, the latter implying maximum baselines of 1 km. The binary fields were produced with the K-Means algorithm as described in section~\ref{sec:ThS}. 

We see that the segmentation of the 21-cm signal and the ionization fraction data at the resolution of the simulation give the same values for $\hat{x}_\mathrm{v}$, hence K-Means recovers the ionized regions well. Even at this resolution the measured value is always lower than the actual one, with differences reaching $\sim 20$ per cent. When creating the binary field, partially ionized cells with ionization fractions below the threshold value will be classified as neutral and do not contribute to the measured global ionization fraction. On the other hand, partially ionized cells above the threshold will contribute 100 per cent to the measured global ionization fraction. The results show that the missing contribution of the former group dominates over the additional contribution of the latter group.

The measured global ionization fraction deviates even more from the actual one after reducing the resolution (dashed and dash-dotted lines in Fig.~\ref{fig:x_obs_kmeans}). While smoothing the data one can on the one hand expect ionized bubbles below a certain size to be no longer visible while on the other hand larger bubbles may appear even larger due to apparent joining. From the reduced values of $\hat{x}_\mathrm{v}$ for lower resolution we infer that the first effect dominates. For the lowest resolution considered here, the measured global ionization fraction can be less than half of the actual value. 

We conclude that it will be hard to obtain an accurate determination of the actual global ionization fraction from tomographic images. Values of ${x}_\mathrm{v}$ and $\hat{x}_\mathrm{v}$ for a number of representative redshifts are given in Table~\ref{tab:z_list}. These are the redshifts for which BSDs are presented in the following section.

Below a global ionization fraction of $x_\mathrm{v}$ $\approx 0.10$, the $\hat{x}_\mathrm{v}$ derived from the 21-cm signals become very noisy. Here the K-Means method has difficulty in dividing the PDF of the signal into ionized and neutral values since the number of ionized data points is very low during the early times. Therefore, K-Means is not a good classifier for the 21-cm signal from the early stages of reionization.

\begin{table}
	\centering
	\caption{A list of the global ionization fractions at different redshifts for our fiducial simulation LB1. $x_\mathrm{v}$ gives the average value of the ionization fraction data cube. $\hat{x}_\mathrm{v,sim}$ gives the fraction of 21-cm cells labelled as ionized after segmentation. $\hat{x}_\mathrm{v,smooth}$ gives the same fraction for 21-cm data cubes smoothed to SKA1-Low resolution.}
	\label{tab:z_list}
	\begin{tabular}{lccccc} 
		\hline
		z & $x_\mathrm{v}$ & $\hat{x}_\mathrm{v,sim}$  & $\hat{x}_\mathrm{v,smooth}$ \\
		\hline
		6.4 & 0.90 & 0.88 & 0.83  \\
        6.7 & 0.70 & 0.64 & 0.49  \\
        6.8 & 0.60 & 0.54 & 0.40  \\
		6.9 & 0.50 & 0.45 & 0.28 \\
		7.3 & 0.30 & 0.24 & 0.12  \\
        7.4 & 0.25 & 0.20 & 0.09  \\
        7.8 & 0.15 & 0.11 & 0.05  \\
		\hline
	\end{tabular}
\end{table}

\begin{figure}
  \centering
  \includegraphics[width=0.5\textwidth]{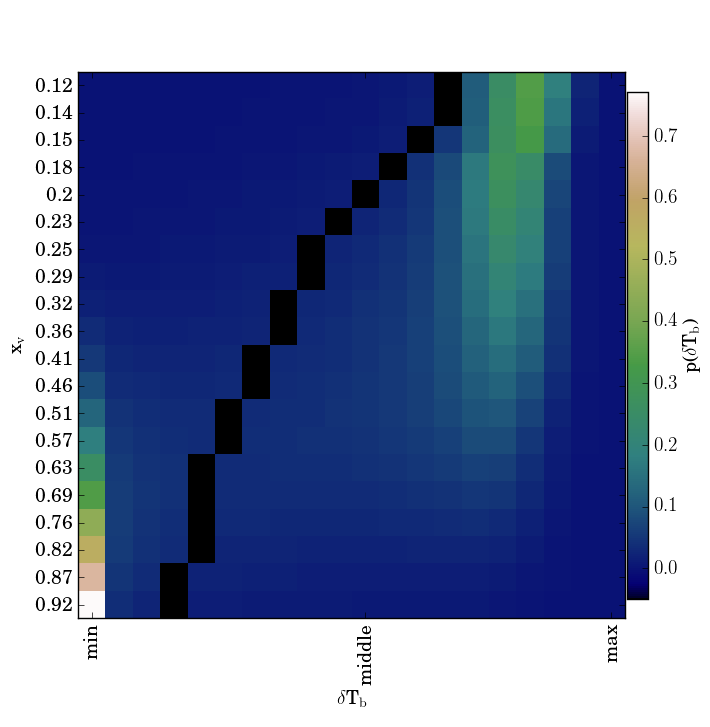}
  \caption{The PDF of observed 21-cm signal 
  at different redshifts shown as a colour map. Each horizontal slice represents a PDF at a particular global ionization fraction (vertical axis). The horizontal axis shows the binned values of $\delta \mathrm{T_b}$ which have been rescaled between their minimum and maximum. {The bin labelled as {\lq middle\rq} is average value of minimum and maximum.} The colours represent the number density of the PDF which is normalized to unity. Along each of the PDFs a black spot indicates the threshold value found by the K-Means algorithm.}
  \label{fig:kmeans_thres}
\end{figure}

To better appreciate the performance of the K-Means algorithm we show in
Fig.~\ref{fig:kmeans_thres} where it places the threshold value with respect to the  PDFs of $\delta T_\mathrm{b}$. The colours show the value of its PDF from early (top) to late stages (bottom) of reionization against the values of  $\delta T_\mathrm{b}$ (scaled from minimum to maximum). The column at the minimum value correspond to the highly ionized cells and the bright areas near $\delta T_\mathrm{{max}}$ correspond to the neutral cells. The spread in the latter is due to the density fluctuations. The threshold should be such that it separates these two modes. The black spots indicate where K-Means puts the threshold value. We can infer that the method works quite well for most of reionization epochs.  
However, for the earliest stages K-Means places the threshold value very close to the density fluctuation mode. As a consequence it identifies points in the tail of the density mode as ionized. 
The ionized cluster that K-Means looks for is so small that the method cannot define a prominent centroid and as a result both centroids are found inside the density fluctuations cluster. This behaviour explains the noisy results in Fig.~\ref{fig:x_obs_kmeans}. For early stages of reionization, a different threshold algorithm should be considered. We postpone a further discussion of this to a future paper (Giri et al., in preparation).

\begin{figure*}
  \includegraphics[width=0.85\textwidth]{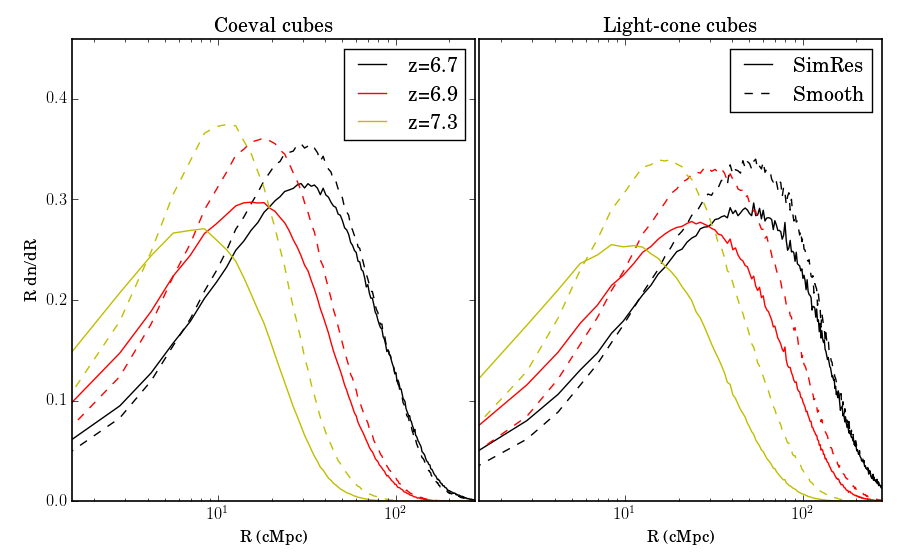}
  \caption{MFP results: The left panel displays the BSDs for CCs at three different redshifts at the resolution of the simulation (dashed) and at SKA1-Low resolution (solid). The right panel display the size distribution of the LC with the indicated central redshift at the resolution of the simulation (dashed) and at SKA1-Low resolution (solid). The BSDs for $z$ = 7.3, 6.9, and 6.7 are shown as the curves from left to right, respectively. The corresponding ionization fractions are given in Table~\ref{tab:z_list}.}
  \label{fig:MFP_2km}
\end{figure*}

\begin{figure*}
  \includegraphics[width=0.85\textwidth]{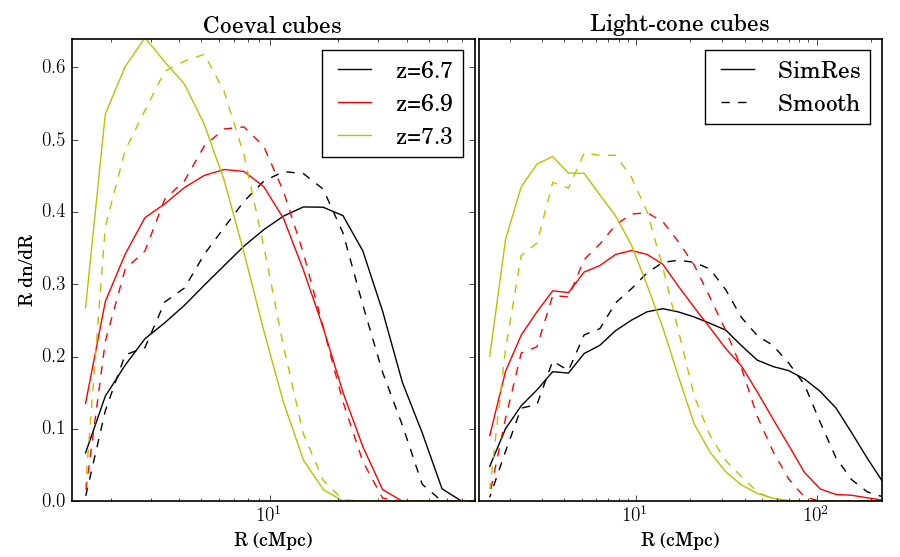}
  \caption{The same as Figure~\ref{fig:MFP_2km} but for the SPA-BSD.}
  \label{fig:SPA_2km}
\end{figure*}

\begin{figure*}
  \includegraphics[width=0.85\textwidth]{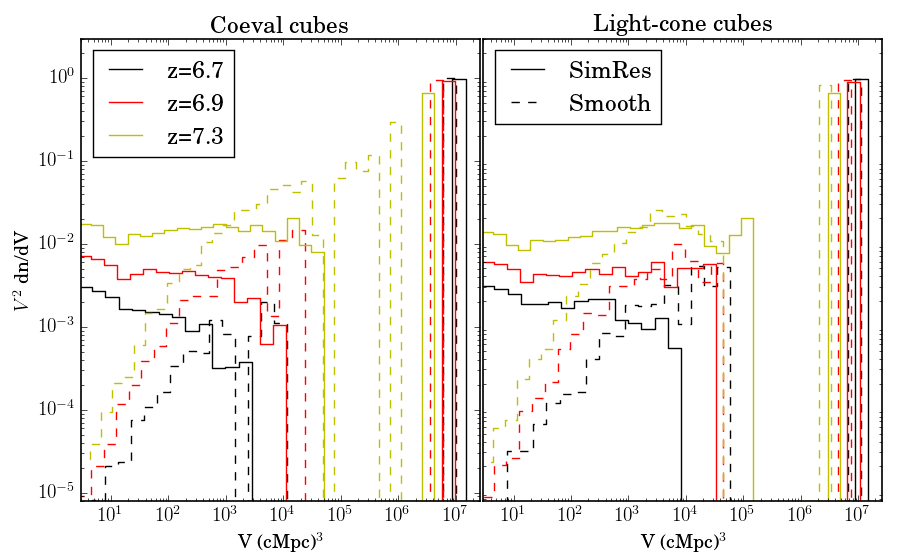}
   \caption{FOF Curves: The left panel shows the volume distribution $V^2{\mathrm d}P/{\mathrm d}V$ of bubbles in CCs at different redshifts at the resolution of the simulation (solid) and at SKA-Low resolution (dashed). The right panel displays the volume distribution of the LC with the indicated central redshift at the resolution of the simulation (dashed) and at SKA1-Low resolution (solid). The BSDs for $z$ = 7.3, 6.9, and 6.7 are shown as the curves from left to right, respectively. The corresponding ionization fraction is given in table~\ref{tab:z_list}.}
  \label{fig:FOF_2km}
\end{figure*}

\subsection{Effect of limited resolution on bubble size distributions}
\label{sec:LimRes}

Now that we have established the performance of the K-Means algorithm and the effect of smoothing and segmentation on the ionization fractions, we can address the performance of the different BSDs introduced in Section~\ref{sec:BS_methods}. In this section, we address the effects of using the 21-cm signal at limited resolution.

In left panels of Fig.~\ref{fig:MFP_2km} \& \ref{fig:SPA_2km}, we compare the SPA- and MFP-BSDs at the resolution of the simulation (solid lines) to the ones at a SKA1-Low resolution (dashed lines). We have picked three redshift values for which the intrinsic and measured global ionization fractions are listed in Table~\ref{tab:z_list}.
Both methods show that {during the early stages of reionization} the peaks of the BSDs shift to larger sizes after reducing the resolution, which is a consequence of smaller bubbles being smoothed out and larger bubbles thus taking up a larger fraction of the distribution. For the later stages ($z=6.7$), 
{we notice that the relative frequency of both the smallest and largest regions is reduced in the smoothed data making the distributions more narrow. This is seen in both the MFP and SPA distributions but is more clear in the latter. As a result the peak value falls at a smaller radius in the smoothed case. As the largest regions display quite complex morphologies with tunnels, bridges and islands, we attribute this behaviour to the smoothing removing some of the connections which exist in the non-smoothed case.}

\citet{2017MNRAS.471.1936K} also noted a shift to larger sizes in the BSDs found by the granulometry method and attributed this to smoothing causing an apparent joining of ionized regions, labelling this effect as the {\lq smoothing bias\rq}. However, normalized curves as the ones we are using here and also used in \citet{2017MNRAS.471.1936K}, display the {\it fraction} of bubbles at a particular size. Therefore a reduction in the absolute number of smaller regions can shift the peak of the distribution to larger values without the need of increasing the absolute number of larger regions. The impact of apparent joining of ionized regions can therefore not be determined from these BSDs. 

Due to their change of shape, the BSDs at lower resolution show less evolution than those for the full resolution case. As a result, it may be hard to distinguish between two
different stages of reionization based solely on these measured curves. However, these normalized BSDs are not sensitive to the total number of ionized regions or the global ionized fraction. To track the progress of reionization, we therefore need to analyze these BSDs jointly with the measured global ionization fractions $\hat{x}_\mathrm{v,smooth}$. As can be seen from  Table~\ref{tab:z_list}, $\hat{x}_\mathrm{v,smooth}$ does evolve substantially, from 0.12 to 0.49, for the low resolution case.

The MFP-BSDs (Fig.~\ref{fig:MFP_2km}) and SPA-BSDs (Fig.~\ref{fig:SPA_2km}) show similar behaviour and the relative shifts of the curves are very similar in both cases. However, it should be noted that the radius at which the BSD peaks is always lower for the SPA method than for the MFP method. For example, the peak distribution values of the SPA-BSD for $\hat{x}_\mathrm{v}=0.1$ and $\hat{x}_\mathrm{v}=0.4$ differ by about 6 Mpc, whereas for the MFP-BSD we see a difference of about 20 Mpc. Hence, we confirm the result from \citet{2011MNRAS.413.1353F, 2016MNRAS.461.3361L} that the peak values of the SPA-BSDs are around three times smaller than the peak values of the MFP-BSDs. \citet{2016MNRAS.461.3361L} have shown that this is due to the strong bias of the SPA method.

\citet{2016MNRAS.461.3361L} have also shown that shape of the MFP-BSD is closer to the intrinsic BSD leading to the MFP method being preferred over the SPA method. Since the MFP method uses random positions and directions for the rays, it can be sensitive to sampling noise, as can be seen in the results for the later stages of reionization in Fig.~\ref{fig:MFP_2km}. This noise can be reduced by increasing the number of rays being traced.

Fig.~\ref{fig:FOF_2km} (left panel) shows the FOF-BSD at the resolution of the simulation (solid) and at SKA1-Low resolution (dashed). As usual, the distribution is bimodal with one large ionized region that 
dominates the total ionized volume and a population of much smaller regions making up the rest. The large ionized region has been referred to as the {\lq percolation cluster\rq} \citep[see ][]{2016MNRAS.457.1813F}, and appears at $x_\mathrm{v} \approx 0.10$. It forms when almost all the ionized regions connect through small bridges and is a distinct feature of any percolation process. The reduction in the amplitude of the mode containing the smaller bubbles illustrates that this population becomes less important as reionization progresses. 

The FOF results clearly show the impact of limited resolution on the population of small regions, as the smallest bubbles are suppressed by a factor of more than 1000. However, we also see a joining effect since in the low-resolution data the largest smaller regions tend to be larger than in the high-resolution case, which suggests that joining does affect the measured BSDs, also for the MFP and SPA methods. Note however, that the percolation cluster actually has a somewhat smaller volume in the low resolution case. For $z=7.3$, the percolation cluster has not yet fully developed in the smoothed data whereas a series of smaller regions in the volume range $10^5$--$10^6$~Mpc$^3$ is detected. This is actually consistent with the value of $\hat{x}_\mathrm{v}$ which is 0.12 for this redshift. The percolation cluster typically emerges around this value \citep{2016MNRAS.457.1813F}. Hence, the reduction in resolution makes the observable smaller bubbles larger and percolation cluster smaller. As the percolation cluster dominates the entire ionized volume, the measured ionization fraction is always lower at lower resolution, consistent with the results in the previous section.

\citet{2016MNRAS.457.1813F} have predicted that the $V^2 \mathrm{d}n/\mathrm{d}V$ curve for the population of smaller regions determined by the FOF method should be flat due to the nature of reionization as a percolation process. 
We indeed observe this behaviour 
at simulation resolution. However, after smoothing the slope becomes positive. Interestingly for all cases, the slope is such that $V^2 dn/dV \propto V$ or $dn/dV \propto V^{-1}$. If this transition to a positive slope is a universal result, FOF-BSDs from observations could still be used to confirm the percolative behaviour of the reionization process.

\begin{figure*}
  \includegraphics[width=0.7\textwidth]{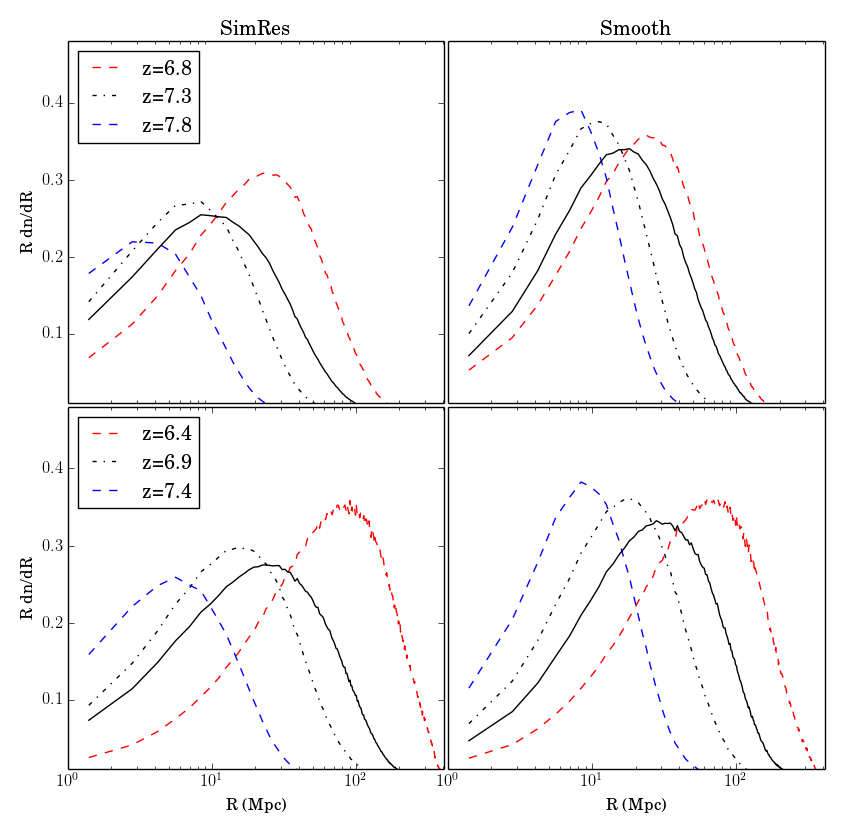}
  \caption{The LC effect in MFP measurements. The BSDs from the two LC data sets (black solid) are compared to the coeval BSDs of the central (black dot-dashed), lowest (blue dashed) and highest (red dashed) redshifts in the LC. The left panels show the results at the resolution of the simulation and the right panels at SKA1-Low resolution.}
  \label{fig:MFP_CC_LC}
\end{figure*}

\begin{figure*}
  \includegraphics[width=0.7\textwidth]{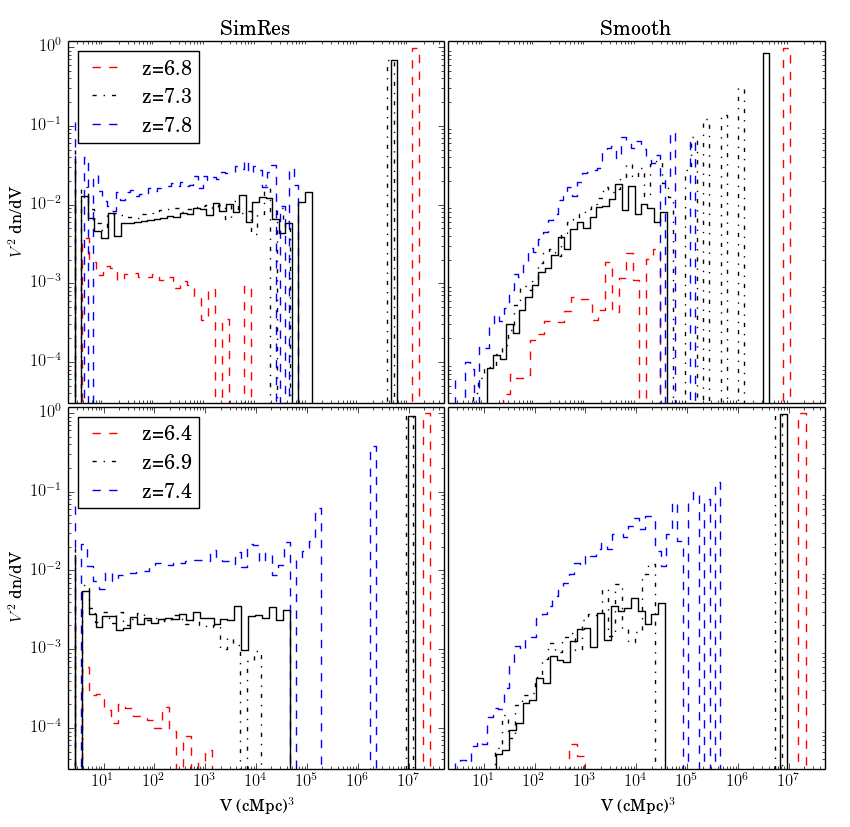}
  \caption{As Figure \ref{fig:MFP_CC_LC} but for FOF-BSDs.}
  \label{fig:FOF_CC_LC}
\end{figure*}

\begin{figure}
  \includegraphics[width=0.45\textwidth]{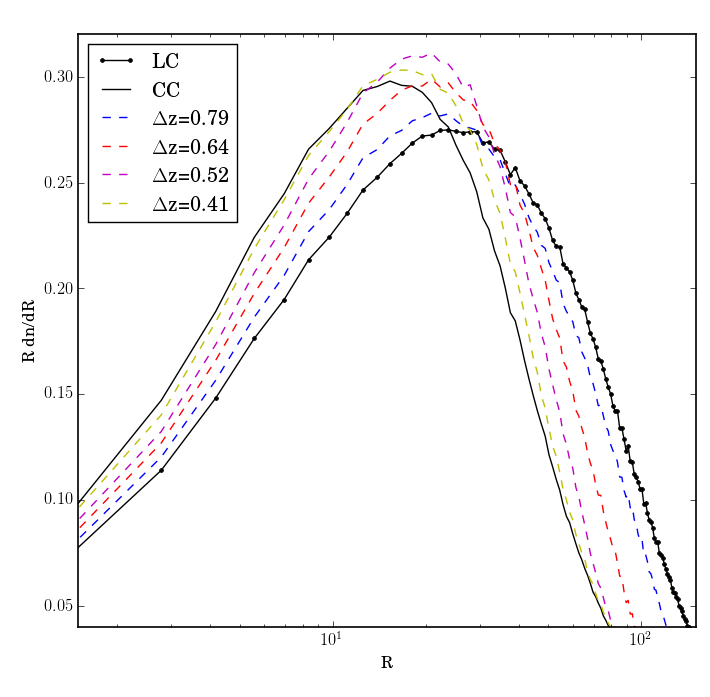}
  \caption{The dependence of the LC effect on the bandwidth used. The black solid curve represents the MFP-BSD for the CC at the central redshift whereas the solid black curve with dots shows the results from the LC from Fig.~\ref{fig:MFP_CC_LC}, bottom left panel ($z=6.9$, $\Delta z=0.82$). The coloured dashed curves show the size distributions for LCs with the same central redshift but narrower bandwidths. The BSDs from the LC data converge to the CC one as the bandwidth is being reduced.}
  \label{fig:MFP_CC_LC_depth}
\end{figure}

\subsection{Line-of-sight evolution}

The previous subsection described the results for CC, but the observations will of course deliver LC image cubes instead, where the frequency axis covers the signals from a range of redshifts. In this section, we consider the impact of the LC effect.


The right panels in Fig.~\ref{fig:MFP_2km} -- \ref{fig:FOF_2km} show the different BSDs for light-cone data of which the central redshift coincides with the redshift values indicated in the figure. The width of the light-cone corresponds to a distance of 349~Mpc which is roughly $\Delta z \approx 0.80$ and is the same as our simulation volume. We see that the LC effect affects all BSDs, pushing them to larger sizes than found in the coeval cubes. The smoothing affects the LC data in a similar way as it does for coeval data. The largest difference is seen for the FOF distribution at early times ($z=7.3$), where the population of larger bubbles that appeared in the coeval data after smoothing is absent in the LC data and the percolation cluster is again apparent. It should however be noted that conclusions for large regions are sensitive to sample variance effects as they are based on only one or two regions.

\citet{2012MNRAS.424.1877D} showed that the BSD determined from LC data can be approximated by the one from the coeval cube at the central redshift of the LC data. They used SPA for their analysis and only considered one redshift value. In Fig.~\ref{fig:MFP_CC_LC}, we compare the MFP-BSD for LC data with the distributions from coeval data corresponding to the highest, central and lowest redshift contained in the LC. The left panels show the MFP curves for the signal at the resolution of the simulation and the right panels the same for SKA1-Low resolution. We see that the MFP-BSD is bracketed between those for the higher and lower redshift coeval cubes, also for the smoothed case. The plot also indicates that the BSD for the central redshift is not a good representation of the one from the LC data. The LC data reveal the presence of larger bubble sizes and its BSD appears to fall in between those from the central and lowest redshifts. The SPA-BSDs (not shown) exhibit a similar behaviour. 

Fig.~\ref{fig:FOF_CC_LC} shows the same analysis for the FOF-BSD. These results present a mixed message. On the one hand, the sizes of the percolation cluster for the LC data is larger than at the central redshift. On the other hand, the distribution of smaller regions in LC case appears to fall in between that seen in the central and highest redshifts, although it is much closer to that of the central redshift.

As studied in more detail in \citet{2014MNRAS.442.1491D}, the impact of the LC effect depends on the width of the LC data set. If the evolution of the signal over the extent of the LC is weak or linear, statistical measurements will be similar to those at its central redshift. However, if there is substantial evolution, this will no longer be the case. \citet{2014MNRAS.442.1491D} recommended that LC data should not have an extension larger than $\Delta z \approx 0.50$ (which during reionization corresponds to a frequency width of $\sim 10$~MHz). The LC data presented above have a width $\Delta z \approx 0.80$.

To explore the effect of the width of the LC, we show BSDs for different LC widths in Fig.~\ref{fig:MFP_CC_LC_depth}. To keep the data sets cubic in the sense that they have the same comoving size in all directions, we select smaller cubes from the large LC cube. We tested that the reduced volumes affect the BSDs in a marginal way. These results confirm the conclusion from \citet{2014MNRAS.442.1491D} that the LC effect becomes a minor nuisance for widths $\Delta z \lesssim 0.50$. 

\subsection{Effect of RSD}
Early studies have shown that RSDs have appreciable impact on the 21-cm power spectrum \citep{2001JApA...22...21B, 2004MNRAS.352..142B}. The matter on average moves towards the high-density regions, therefore in redshift space RSDs tend to compress high density regions and to expand low density regions. This is known as the Kaiser effect \citep{1987MNRAS.227....1K}. Clearly, the sizes of the bubbles observed using the redshifted 21-cm signal could also be affected by the gas peculiar velocities. As the ionized regions are typically associated with the high density, source rich, regions, we expect that RSDs will decrease the observed bubble sizes. As shown by \citet{2013MNRAS.435..460J}, the effect of RSDs is largest during early reionization and becomes progressively weaker as reionization progresses. 
{Close inspection of Fig.~\ref{fig:LC_SmoothLC} indeed reveals small but observable differences in the shapes of ionized regions between the non-distorted and distorted cases, even at SKA1-Low resolution.}

\begin{figure}
  \includegraphics[width=0.45\textwidth]{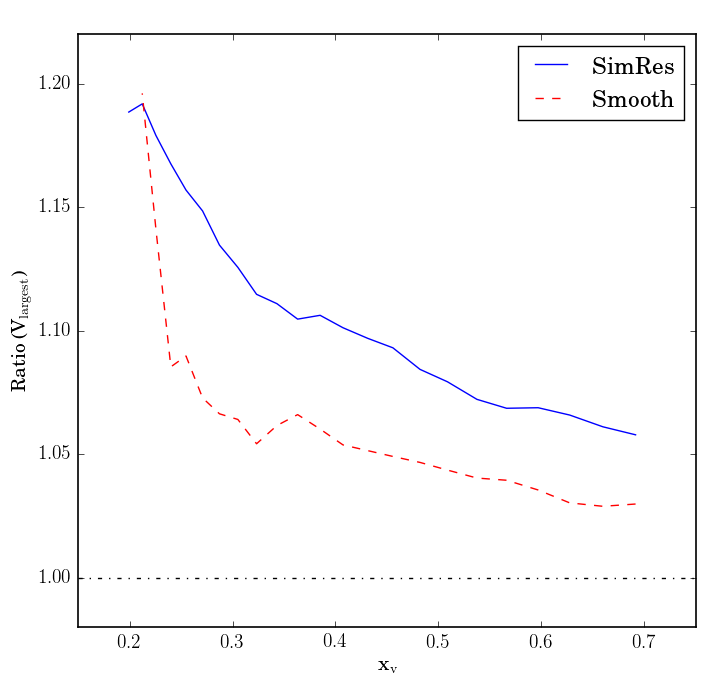}
  \caption{The ratio of the volume of the largest ionized region found in the sub-volume from the light-cone without RSD to the ones with RSD vs. the global ionization fraction ($x_\mathrm{v}$). The largest region is found using the FOF algorithm.}
  \label{fig:RSD_ratio}
\end{figure}

\begin{figure}
  \includegraphics[width=0.45\textwidth]{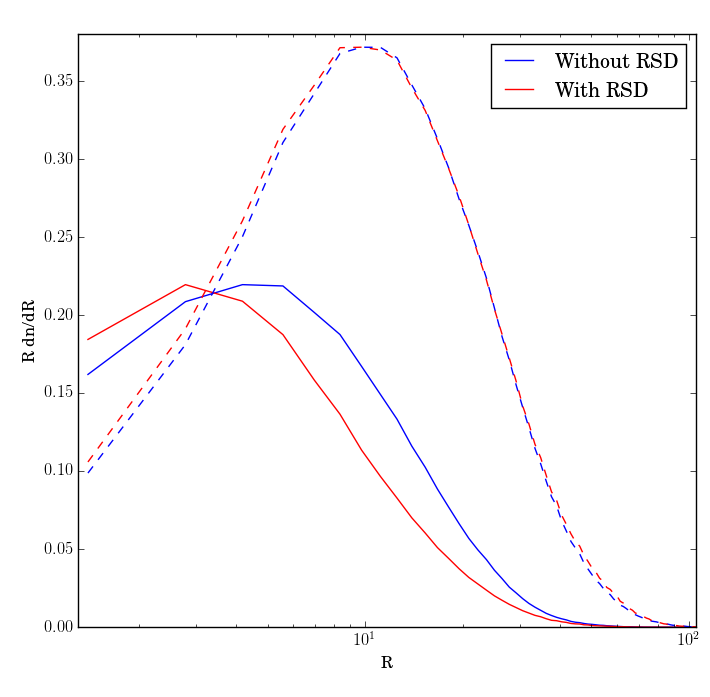}
  \caption{The MFP-BSD of sub-volume from light-cone with RSD and without RSD are given. The solid curve gives ones at the simulation resolution and teh dashed ones are for the lower resolution case. All the plots are for the epochs with $x_\mathrm{v}=0.15$.}
  \label{fig:MFP_smLC_smRSD}
\end{figure}

To study the effect of RSDs on the BSDs, we first consider the volume of the largest connected region in the cube as found by the FOF method at different ionization fractions $x_v$. Fig.~\ref{fig:RSD_ratio} shows the ratio of this volume from a light-cone cube (width 349~Mpc) without RSD to one with RSD. We consider both the the simulation resolution (solid curve) and SKA1-Low resolution (dashed curve). We see that this largest region is larger without RSD and that the size ratio approaches unity as reionization progresses. This result confirms our expectation that the RSD effect decreases the measured bubble sizes. However, the results also show that the magnitude of this effect is at most 20 per cent and for most of reionization even lower.

In Fig.~\ref{fig:MFP_smLC_smRSD} we show the effect of RSDs on the MFP-BSD at $z=7.8$. The redshift choice is owing to the previous inference that RSDs have a larger effect earlier during reionization. We see a shift to the smaller $R$ in the BSDs for the RSD case. This again supports the idea that RSDs decreases the observed sizes. However, the two BSDs at SKA1-Low resolution (dashed lines) are almost identical, which indicates that even though RSDs have an effect on the sizes, this may not be detectable in  low-resolution data.

Both of these results show that RSDs do not have a major impact on the size distribution of the ionized regions from the observed low-resolution data. This can be understood from the realization that RSD affect the sizes of the ionized region in only one direction (along the frequency direction). Since the MFP, SPA and FOF methods all consider three-dimensional structures, the small change in size along one dimension caused by the RSDs is mostly averaged away.

\subsection{Comparing different source models}
One of the most important reasons to the study the 21-cm signal from the EoR is to understand the nature of the sources of reionization. Hence, the variation in the BSDs for different source models and whether BSDs can be used to differentiate them are relevant to study. Although a full exploration is beyond the scope of this paper, we compare here BSDs for three different source models taken from \citet{2016MNRAS.456.3011D}, namely simulations LB1 (only massive sources), LB3 (partial suppression of low-mass sources) and LB4 (gradual, mass-dependent suppression of low-mass sources).

\begin{figure*}
  \includegraphics[width=0.9\textwidth]{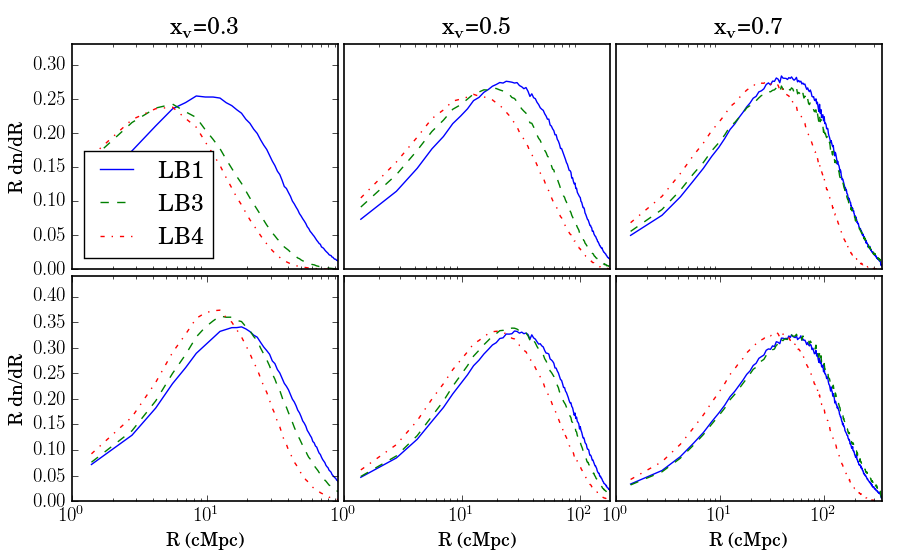}
  \caption{MFP-BSDs for the three simulations considered. We compare them at epochs corresponding to $x_\mathrm{v}=0.3,0.5,0.7$. The upper panel shows the comparison at the resolution of the simulation whereas the lower panel shows the study of smoothed signal.}
  \label{fig:MFP_model_compare}
\end{figure*}

In Fig.~\ref{fig:MFP_model_compare}, we show the MFP-BSDs for these three source models at different epochs ($x_\mathrm{v}=0.3,0.5,0.7$) at the resolution of the simulation (upper panels) and at the SKA1-Low resolution (bottom). We see that initially case LB1 is quite different from LB3 and LB4, since it does not have low-mass sources, resulting in later reionization with larger ionized bubbles. However, as reionization progresses the MFP-BSDs of LB1 and LB3 become more similar since low-mass sources become partially suppressed over time, resulting in late-time reionization being dominated by the same high-mass sources in both cases (although the timing of reionization is quite different in the two cases). In the LB3 case, the suppression of low-mass sources is mass-dependent, so the lowest-mass ones are most suppressed, while larger ones remain less affected. This yields a different BSD, shifted towards somewhat smaller scales at any given stage of the reionization history.  

At SKA1-Low resolution, the MFP-BSDs for cases LB3 and LB4 are initially more different than in the unsmoothed case, due to the difference in their suppression mechanisms and the different timing of reionization. Otherwise we see the same behaviour, except that, as already noted above, the evolution of the curves spans a smaller range in $R$ values. Given that the horizontal axes in these panels are logarithmic, these curves should be clearly distinguishable when observed at high enough signal to noise to identify the ionized regions.

We performed a similar analysis as in Fig.~\ref{fig:MFP_model_compare}, but now based on the FOF-BSDs (figure not shown). The largest differences in the results are in the volume of the largest connected region and in how large the largest of the population of smaller regions are. 
However, the differences appear to be small and the FOF-BSDs do not appear to be a good tool to distinguish different source models. The most likely cause is that the form of FOF-BSD is dominated by the nature of reionization as a percolation process \citep{2016MNRAS.457.1813F}, with modest dependence on the details of the different models, although a confirmation of this conclusion requires analysing a larger set of simulations.

\citet{2016MNRAS.456.3011D} compared the 21-cm power spectra of these same three models and found that they also differ. This implies that for these three cases the BSDs do not break a degeneracy which could be present in a power spectrum analysis. 
However, the differences in the power spectra are only in the amplitude of the curve when we consider the observable range of $k$ values ($k \lesssim 0.5$~Mpc$^{-1}$). The shapes and slopes of the curves are quite similar and the power spectra do not show clear features at a specific scale. This makes the power spectra sensitive to calibration errors in the absolute flux scale or to foreground of instrumental residuals which add power to the signal (Cathryn Trott, priv.\ comm.). The BSDs are insensitive to deviations in the flux scale and could therefore be used to reduce the uncertainties while comparing the observations to models.

\subsection{Percolation}

\begin{figure}
  \includegraphics[width=0.45\textwidth]{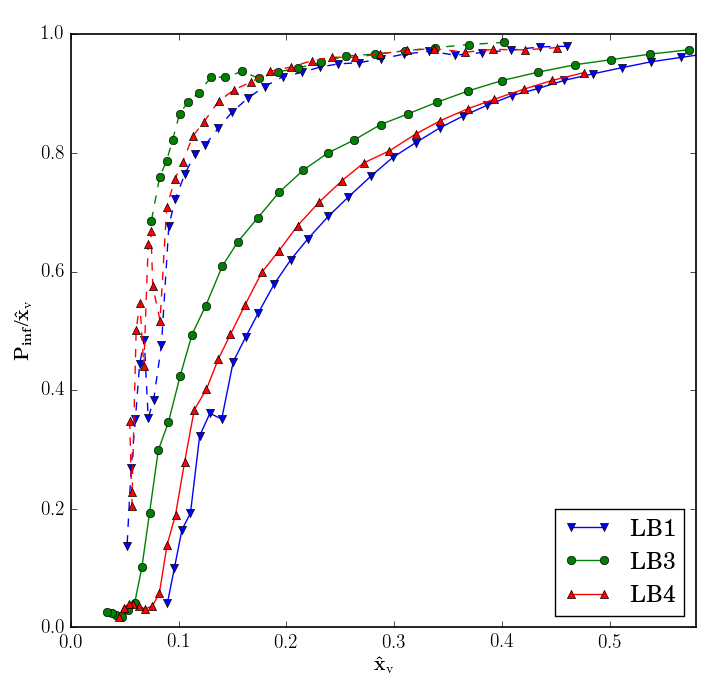}
  \caption{The relative volume of the largest cluster identified by the FOF method plotted against the measured global ionization fraction $\hat{x}_\mathrm{v}$. The solid lines show the results at the resolution of the simulation and the dashed curves at SKA1-Low resolution. The three colours show the results for the three different simulations.}
  \label{fig:model_percolation}
\end{figure}

We mentioned above that the FOF-BSDs do not distinguish clearly between the three source models. However, there is an another aspect to the FOF method, which is the emergence of the dominating largest ionized region. The growth curves for this percolation cluster were studied by \citet{2016MNRAS.457.1813F}. They found that for a range of models the rapid rise in the growth curve happens around $x_\mathrm{v} \approx 0.1$ and that this behaviour is expected from percolation theory.

In Fig.~\ref{fig:model_percolation}, we show growth curves of the largest region found by the FOF method for the three source models considered. The fraction of the ionized volume contained in the largest connected region is plotted against the measured ionization fraction. As above, we consider both the resolution of the simulation (solid curves) and SKA1-Low resolution (dashed curves).

These percolation curves show the same behaviour as noted by \citet{2016MNRAS.457.1813F}; around $\hat{x}_\mathrm{v} \approx 0.1$, the largest cluster starts a rapid growth and contains most of the ionized volume before $\hat{x}_\mathrm{v} \approx 0.2$ is reached. We see some differences between the curves of the different models with LB3 showing the earliest and most rapid growth. This can be understood from the presence of a large population of low mass sources in that model, which also shifts the evolution to earlier times.

For the lower resolution results (dashed curves) the results are more similar between the three models and the rise starts earlier, around $\hat{x}_\mathrm{v} \approx 0.06$. It is also much more rapid, reaching a fraction of 80 per cent around $\hat{x}_\mathrm{v} \approx 0.1$. The reduced resolution thus leads to increased connectivity and a larger relative size for the percolation cluster at a given observed mean ionized fraction. 

In section~\ref{sec:LimRes}, we saw that smoothing decreases the size of the observed percolation cluster. However, lower resolution causes the $\hat{x}_\mathrm{v}$ to decrease as well. As the slope of the curve is greater than unity, the decrease in both numerator and denominator causes the fractional volume to increase. As shown in Sect.~\ref{sec:global_ionfrac} and Table~\ref{tab:z_list}, the observed mean ionized fraction underestimates the mean ionized fraction from the simulation, implying that the observed transition takes place around ${x}_\mathrm{v} \approx 0.2$.

The shape of these percolation curves is sensitive to the chosen threshold value for segmenting the data into ionized and neutral clusters. \citet{2016MNRAS.457.1813F} used a threshold value for the ionized fraction that made the total fractional volume of ionized regions equal to the mean ionized fraction at a given redshift. Since this quantity is unknown for real observations, we used thresholds values for the 21-cm signal determined by the K-Means method, which means we cannot compare in detail to the results in \citet{2016MNRAS.457.1813F}. We should also point out that for the low mean ionization fractions at which the transition happens ($\hat{x}_\mathrm{v} \lesssim 0.1$), K-Means has difficulty finding good values for the threshold, leading to some noisiness in the results.

\section{Summary and Discussion}

In this work, we propose a new approach to analyse tomographic 21-cm data sets from reionization. 
It consists of two steps: first, the 21-cm data is segmented into a binary field of ionized and neutral regions. 
After that, one of the existing BSD methods
can be applied to this field. For the first step we introduced a new method, known as K-Means clustering. For the second step, we have investigated three different BSD methods 
- mean free path (MFP), spherical average (SPA) and friends-of-friends (FOF), each with its own strengths and weaknesses. In particular, we are interested in how they perform when applied to 21-cm data cubes generated by future radio interferometers such as SKA1-Low. 

We considered a number of effects which will be present in the 21-cm data cubes:
\begin{enumerate}
\item Finite resolution (corresponding to maximum baselines of 2~km)
\item Absence of zero base lines (causing the average signal in an image to be zero)
\item Light-cone effect
\item Redshift space distortions
\end{enumerate}
We did not consider the effects of noise and telescope calibration.

The K-Means algorithm can be described as a self-adjusting thresholding technique. Use of such a technique is important in view of the reduced resolution and lack of an absolute zero point in the interferometric observation. We find K-Means to work well if a sufficient number of ionized resolution elements are present in the data. In terms of the measured volume-averaged ionization fraction of the IGM, we find this criterion to imply $\hat{x}_\mathrm{v} \gtrsim 0.1$. For lower values of $\hat{x}_\mathrm{v}$, other methods may perform better. However, it is also possible that at these early times it is fundamentally difficult to distinguish between small, partly resolved ionized regions and density fluctuations.

The results from the different BSD methods show some shared trends, while also describing different aspects of the ionization field.  
Of the three BSD methods we considered, MFP proves to be somewhat more useful than SPA, largely because it is less diffusive. We confirm the result of \citet{2011MNRAS.413.1353F} and \cite{2016MNRAS.461.3361L} that the bubble sizes from the SPA- and MFP-BSDs are not directly comparable due to their different biases; the SPA method gives roughly three times smaller bubbles than the MFP method. The FOF results cannot be compared to either the MFP or SPA methods, since unlike those methods it is finding the volumes of topologically-connected regions. Based on our limited exploration, FOF appears to be mostly useful to confirm the nature of reionization as a percolation process, but does not clearly distinguish between models with different properties for the sources of reionization.

When the resolution is decreased, the BSD curves from the SPA and MFP methods at early times show a shift to larger sizes. This shift diminishes with the progress of reionization and at the later stages of reionization it becomes marginal. Consequently, the MFP- and SPA-BSD curves at the typical resolution of SKA1-Low show less evolution than the intrinsic ones at simulation resolution. In the presence of errors and sample variance the derivation of the reionization history solely from these BSDs may therefore be difficult. However, taking into account the global ionized fraction measured from the binary data set may be able to alleviate this difficulty.

The FOF-BSDs show that the largest (or percolation) cluster is always smaller for the lower resolution data, which may explain why the shift seen in the MFP- and SPA-BSDs decreases as reionization progresses. Late in reionization the size measurements by the MFP and SPA algorithms will mostly take place inside the large percolation cluster. In fact, both the SPA and MFP results show a hint of having a lower fraction of regions at the largest sizes.

Real 21-cm data sets will be affected by both the LC effect and RSDs. We found that the impact of the LC effect is only significant if the data extend for more than $\sim 10$~MHz in frequency or about 0.5 in redshift. The RSDs typically change the BSDs by at most 10 per cent at the simulation resolution and much less at SKA1-Low resolution and can therefore be safely ignored when measuring and interpreting BSDs. 

Due to the strongly non-Gaussian PDF of the 21-cm signal, a power spectrum analysis in principle may suffer from degeneracy as it does not provide a complete statistical description of the results. BSDs obtained from tomographic imaging data should be able to break these degeneracies. However, for the three models studied in this paper, both the power spectra and BSDs are different and therefore it remains to be shown that BSDs can distinguish scenarios which show very similar power spectra. Still, even if such scenarios never occur in reality, the measured BSDs will be affected differently by measurement and calibration errors and will therefore improve the reliability of the astrophysical and cosmological parameters derived from the 21-cm data.

In this study, we have assumed that the noise level in the tomographic data is low enough not to affect the segmentation and the BSD measurements. However, as for example shown in \citet{2017MNRAS.471.1936K} this assumption is optimistic since rms noise levels as low as $\sim 2$~mK can impact the results.
A full evaluation of the impact of noise is beyond the scope of the current paper. However, some general considerations can be made. The presence of noise in the signal would first of all affect the segmentation step. If the segmentation procedure labels noisy pixels in the wrong way, erroneous neutral spots might for example appear inside the ionized regions, or vice versa. The SPA method can be expected to be less affected by such erroneous spots, as it determines the size of an ionized region by its average volume filling factor. However, the appearance of erroneous ionized spots may boost the number of small bubbles found. The MFP method may be more sensitive to the presence of erroneous neutral spots, as hitting such a spot with a ray will always reduce the length of the ray compared to what it should be. On the other hand, the MFP method can be expected to be less sensitive to the appearance of erroneous ionized spots, as the random selection process makes the selection of these points as starting points for rays very unlikely. The FOF method would determine volumes that are reduced by the number of neutral spots and also show an increase in the number small ionized volumes due to the erroneous ionized spots. \citet{2017MNRAS.471.1936K} showed that for the  granulometric method noise introduces a {\lq splitting bias\rq} meaning that it shifts the BSDs to smaller sizes by splitting connected regions into separate ones. The same effect can be expected for the FOF method.

The rms noise level not only depends on the integration time, but also on the resolution chosen. The analysis of the tomographic 21-cm data will require a careful balance between a low enough resolution to achieve acceptable noise levels and a high enough resolution to extract useful BSDs. Although we have not presented a detailed resolution study, the lower panel of Fig.~\ref{fig:MFP_model_compare} indicates that reducing the resolution reduces the differences between the BSDs at different phases of reionization and the differences between different models \citep[see also][]{2017MNRAS.471.1936K}. If the telescope data require too low resolution to obtain meaningful BSDs, other statistical measures for the tomographic data should be considered. 

{As explained in Section~\ref{sec:21cm_signal} we assumed the high spin temperature limit when constructing the 21-cm signal. In this case the lowest value for the signal is zero which corresponds directly to the ionized regions. This allows us to identify ionized regions from the 21-cm signal. Although this limit is generally thought to be valid during most of the EoR, it is possible that spin temperature fluctuations exist during reionization. If regions exist with $T_\mathrm{S} < T_\mathrm{CMB}$, the lowest value for the 21-cm signal will be less than zero. It immediately follows that in this case it will be hard or even impossible to identify ionized regions, especially without a calibration of the absolute value of the 21-cm signal. Even if we would know which regions have $\delta T_\mathrm{b}\approx 0$, we would still not be able to tell whether they correspond to regions  with $x_\mathrm{HI}\approx 0$ or $T_\mathrm{S}\approx T_\mathrm{CMB}$. Furthermore, the 21-cm PDFs from models with spin temperature fluctuations have smooth shapes which means it will be hard to define physically motivated threshold values for any type of size analysis.

However, it has also been shown that during the period of spin temperature fluctuations the signal also is significantly non-gaussian \citep{2015MNRAS.454.1416W, 2017MNRAS.468.3785R}
, so it should
be worthwhile to explore tomographic techniques for this regime. However, at this time it is difficult to say whether BSDs, with some other definition of what constitutes a bubble than we have used, are a useful tool in this context or whether other techniques are preferable. This analysis of tomographic data with spin temperature fluctuations needs to be considered carefully.}

In this paper, we considered the three classical methods developed to characterise BSDs in simulation data and applied them to mock 21-cm observations. Recently, two alternative methods for deriving BSDs have been proposed. The first is the watershed method, which was proposed for reionization studies by \citet{2016MNRAS.461.3361L} who used it on simulated $x_\mathrm{HII}$ data and compared to the SPA, MFP and FOF methods. The method has a marker based algorithm \citep{barnes2014priority}. The markers are the points from which `flooding' starts until the watershed lines are reached. \citet{2016MNRAS.461.3361L} use the local minima in the field of distance to the nearest neutral resolution element to determine the markers, which leads to an over-segmentation. The over-segmentation can be controlled by carefully choosing a smoothing parameter. In the comparison, watershed performs well although the authors did `tune' the method to optimize its performance. Its application to  21-cm tomographic data merits further exploration. However, given the results of \citet{2016MNRAS.461.3361L}, we expect the method to show a similar behaviour as we found for the MFP method.

The other new method is granulometry, described in detail in \citet{2017MNRAS.471.1936K}. These authors not only introduced the method, but also considered many of the issues related to applying this method to finite resolution and noisy 21-cm data. In our paper, we have referred in several places to those results. The method shows good promise, but a comparison to the methods used in this paper as well as the watershed method would be useful.

Both the watershed and granulometry method require a segmentation of the data into neutral and ionized elements. \citet{2017MNRAS.471.1936K} chose a very simple approach, namely labelling all regions lower than the mean 21-cm signal as ionized. As explained in more detail in that paper, this choice only properly identifies ionized regions during part of reionization and may erroneously label low density regions as ionized. The granulometry method would clearly benefit from a more robust segmentation method, for example the one used in the current paper.

We considered BSDs of ionized regions. As reionization approaches completion, the concept of discrete ionized regions becomes less and less applicable. We therefore did not consider BSDs beyond global ionization fractions of 0.7. Beyond that a study of the BSDs of neutral regions will make more sense. We did not address this, but the methodology would be completely equivalent and we postpone an investigation of this to a future paper.

BSDs, irrespective of what method or component is chosen, are of course not the only metric that can be applied to tomographic data. Other possible metrics are the Minkowski functionals that describe the global topological characteristics of ionized or neutral regions \citep[see for example][]{2011MNRAS.413.1353F}, metrics which describe the shapes of ionized/neutral regions or metrics which depend on the (relative) positions of ionized/neutral regions. Some of these metrics may be relatively insensitive to the parameters of reionization, but others may be able to break degeneracies in an analysis based solely on power spectra. Metrics for shapes and positions have been previously applied to voids in the galaxy surveys, see for example \citet{2009ApJ...699.1252F}.

This paper presents the first exploration of the application of BSDs on 21-cm tomographic data. As discussed above, several directions for future investigations present themselves. The most important one is the presence of noise, which will mostly complicate the segmentation of the data into ionized and neutral regions. The choice of segmentation method may actually be important to minimize the impact of noise. We will address these issues in a follow up paper. A comparison between the MFP, granulometry and watershed BSDs would be another useful step in finding an optimal size distribution tool for 21-cm tomography. However, we expect the qualitative conclusions from our current work to hold also for these other BSD methods.

\section*{Acknowledgements}

This work was supported by the Science and Technology Facilities Council [grant numbers ST/I000976/1 and ST/P000252/1] and the Southeast Physics Network (SEPNet). GM is supported in part by Swedish Research Council grants 2012-4144 and 2016-03581. We acknowledge that the results in this paper have been achieved using the PRACE Research Infrastructure resources Curie based at the Très Grand Centre de Calcul (TGCC) operated by CEA near Paris, France and Marenostrum based in the Barcelona Supercomputing Center, Spain. Time on these resources was awarded by PRACE under PRACE4LOFAR grants 2012061089 and 2014102339 as well as under the Multi-scale Reionization grants 2014102281 and 2015122822. Some of the numerical computations were done on the Apollo cluster at The University of Sussex. We thank Eiichiro Komatsu and Cathryn Trott for useful discussions.





\bibliographystyle{mnras}
\bibliography{references}



\bsp	
\label{lastpage}
\end{document}